%% file: bb1.tex
\documentclass[onecolumn]{aa}
\usepackage{txfonts}
\usepackage[comma,authoryear]{natbib}
\usepackage[latin1]{inputenc}

\usepackage[dvips]{graphicx}

\input com31

\begin{document}
\author{M. Rieutord}
\institute{
Laboratoire d'Astrophysique de l'observatoire Midi-Pyrénées, UMR 5572,
CNRS et Université Paul Sabatier, 14 avenue E. Belin, 31400 Toulouse,
France}

\title{The dynamics of the radiative envelope of rapidly rotating stars}
\subtitle{I. A spherical Boussinesq model}
\date{\today}
\abstract{The observations of rapidly rotating stars are increasingly
detailed and precise thanks to interferometry and asteroseismology;
two-dimensional models taking into account the hydrodynamics
of these stars are very much needed.}{A model for studying the dynamics of
baroclinic stellar envelope is presented.}{This models treats the
stellar fluid at the Boussinesq approximation and assumes that it is
contained in a rigid spherical domain. The temperature field along with
the rotation of the system generate the baroclinic flow.}{We manage to
give an analytical solution to the asymptotic problem at small Ekman and
Prandtl numbers. We show that, provided the \BV\ frequency profile is
smooth enough, differential rotation of a stably stratified envelope
takes the form a fast rotating pole and a slow equator while it
is the opposite in a convective envelope. We also show that at low
Prandtl numbers and without $\mu$-barriers, the jump in viscosity
at the core-envelope boundary generates a shear layer staying along
the tangential cylinder of the core. Its role in mixing processes is
discussed.}{Such a model provides an interesting tool for investigating
the fluid dynamics of rotating stars in particular for the study of
the various instabilities affecting baroclinic flows or, even more,
of a dynamo effect.}

\maketitle

\section{Introduction}

Thanks to the development of new observational techniques, like
interferometry or high precision photometry, rapidly rotating stars are
more and more focusing the interest of the scientific community. The
best example is certainly the nearby star Altair whose shape, rotation,
inclination of axis have been determined by interferometry
\cite[][]{DKJVONA05,petersonetal05} and which has also been identified as an
oscillating $\delta$-Scuti star \cite[see][]{buzasietal05}.
Modelling such stars is therefore a challenge which needs to be taken up
in order to extract the best science from these observations.

The role of rotation in the physics of stars appears at various levels.
First is the determination of the shape and thermal structure which
control the way we ``see" the star and at this stage the differential
rotation is certainly important. Then, if we are concerned by the
eigenspectrum, in addition to the shape (which influences the frequency
and the visibility
of the modes, see \cite{LRR05}), we also need to know the distribution
of elements able to excite eigenmodes through $\kappa$-mechanism.
Finally, as it has long been known \cite[e.g.][]{SZ70}, rotation may
drive many hydrodynamical instabilities in the stably stratified
radiative zones of rotating stars which inevitably lead to some
small-scale turbulence. Much work has already been done in this
direction following \cite{zahn92}, in particular to understand the
surface abundances in stars across the HR diagram \cite[see][]{MM00}.

The case of rotation is therefore fundamental in stellar physics and
needs to be well understood. Presently, the difficulties concentrate on
the hydrodynamical effects generated by the associated inertial
accelerations (Coriolis and centrifugal ones). Indeed, the analysis of
these flows is quite demanding as it requires two-dimensional models.

The present paper focuses on one side of the general problem of the
dynamics of rotating stars: namely, we want to specify the shape
of the baroclinic flow inside the radiative envelope of a rapidly rotating
star. In such stars, it is well known that the Eddington-Sweet time
scale is short enough for a steady state to be reached. But this steady
flow, in which viscous effects need to be taken into account, has never
been computed in its full two-dimensionality.

Previous work on this subject is quite scarce. First attempts include
those of \cite{TT82,TT83a,TT83b,TT84,TT86} who, for analytical
tractability,  restricted the velocity
field to its first Legendre polynomial component and therefore give an
almost one dimensional description of the flow. Still in the
quasi-one-dimensional approach, the work of \cite{zahn92} presents a
self-consistent description based on the assumption of a strong
horizontal turbulent viscosity compared to the vertical one.
Later, \cite{UE94,UE95} touched upon the question of modeling rotating
baroclinic stars in two dimensions but their neglect of viscosity removed
any meridian circulation and left differential rotation incomplete.
Including viscosity, \cite{garaud02} investigated the dynamics of the
radiative core of the sun stressed by the differential rotation of the
convective zone; in this latter case the flow is essentially driven by
the boundary conditions. This is certainly the
most advanced work on stellar baroclinic flows but baroclinicity is not
the main driving.

Although the full solutions were still not known, the stability of some
generic situation of baroclinic shear flows has been investigated by
\cite{KS82,KS83,knobloch85,schneider90} and are reviewed in \cite{zahn93}.

As one may easily imagine, computing the baroclinic flow in a
centrifugally distorted envelope with realistic density and temperature
profiles is not an easy task. Therefore, the step taken here concentrates
on a simpler version of such a flow, namely a ``Boussinesq version in
spherical geometry'', which, as we shall see, still contains a good deal of
the actual physics of the problem. This model has two purposes: first
to reveal the main dynamical features of such flows and eventually find
some robust property that may persist in actual stars; second to devise
a model in which hydrodynamics can be investigated more easily and which
can be used as a template for the construction of more elaborated models.

The next section of the paper describes our approach, especially the
simplifications that we use or the physics we keep. In section 3,
we discuss the asymptotic properties of this model in the limit of
small Ekman numbers, appropriate to stellar applications. Then, after
describing the numerical method, we investigate some aspects of the
general case. Conclusions and outlooks follow.

\section{The Model}

\subsection{Description}

We consider a system in which a self-gravitating fluid of
constant density is enclosed in an undeformable sphere of radius $R$.
The gravitational field is thus simply $\vg=-g\vr$ where the radial
coordinate is scaled with $R$ (i.e. $0\leq r \leq 1$) and $g$ is the surface
gravity. The thermal and mechanical equilibrium of this fluid is
governed by:

\greq
-\na P_{\rm eq} + \rho_{\rm eq}\vg =\vzero\\
\Div(\khi\na T_{\rm eq}) + Q =0 \\
\rho_{\rm eq}=\rho_0(1-\alpha(T_{\rm eq}-T_0))
\egreqn{equil}
where $\alpha$ is the dilation coefficient, $\khi$ the thermal
conductivity and $Q$ heat sinks (since we study a stably stratified
situation). The essential characteristic of this  equilibrium is the
\BVF\ profile

\beq N^2(r) = \alpha \dnr{T_{\rm eq}} g(r) \eeqn{bvfp}

We now let this system rotating at the angular velocity $\Omega$ around
the z-axis.  In the rotating frame, because of the combination of rotation and
stratification, a steady  baroclinic flow appears.  It is a solution of

\begin{eqnarray*}
\rho(2\vO\wedge\vv + \vv\cdot\na\vv) &=& -\na P +
\rho(\vg+\Omega^2s\es)+\mu\Delta\vv \\
\rho c_p\vv\cdot\na T &=& \Div(\khi\na T) + Q \\
\end{eqnarray*}
where $\mu$ is the dynamical shear viscosity, $c_p$ the specific
heat capacity at constant pressure, $s$ the radial cylindrical coordinate
and $\es$ the associated unit vector.
We decompose the thermodynamical quantities into their equilibrium
and fluctuating parts, namely

\[ P=P_{\rm eq} + \delta P, \qquad \rho=\rho_{\rm eq} + \delta\rho,
\qquad T=T_{\rm eq}+\delta T\]
After subtracting \eq{equil}, the momentum and heat equations read:

\begin{eqnarray*}
\rho(2\vO\wedge\vv + \vv\cdot\na\vv) &=& -\na\delta P +
\delta\rho(\vg+\Omega^2s\es) + \rho_{\rm eq}\Omega^2s\es+\mu\Delta\vv \\
\rho c_p(\vv\cdot\na T_{\rm eq}+\vv\cdot\na \delta T &=& \Div(\khi\na
\delta T)
\end{eqnarray*}
These equations are further simplified by using the \BA\ which yields:

\begin{eqnarray*}
2\vO\wedge\vv + \vv\cdot\na\vv  &=& -\na\delta P - \alpha\delta
T((\vg+\Omega^2s\es) - \alpha\delta T_{\rm eq}\Omega^2s\es +
\nu\Delta\vv \\ 
\vv\cdot\na T_{\rm eq}+\vv\cdot\na \delta T &=& \kappa\Delta\delta T
\end{eqnarray*}
At this stage we need to point out that we removed the barotropic
contribution of the centrifugal acceleration, $\rho_0\Omega^2s\es$, and
that fluctuations of density are retained whenever they are multiplied
by an acceleration ($\vg$ or $\Omega^2s\es$) as should be done when
using the \BA. {\bf We note that this approximation implies the Cowling
approximation (i.e. the neglect of variations of self-gravity).}
We also simplified the system by assuming a constant heat
conductivity. Taking the curl of the momentum equation we derive the
equation of vorticity:

\beq 
\na\times\lp 2\vO\wedge\vv + \vv\cdot\na\vv + \alpha\delta
T((\vg+\Omega^2s\es) - \nu\Delta\vv \rp = -\eps N^2(r)\sth\cth\ephi
\eeq
where we used the \BVF\ profile \eq{bvfp} and introduced $\eps=\Omega^2R/g$, the
ratio of centrifugal acceleration to gravity. These equations are of
course completed by the equation of mass conservation which reads

\[ \Div\vv=0\]
at the \BA.

The foregoing equations show that our problem is that of a forced flow
driven by the baroclinic torque $-\eps N^2(r)\sth\cth\ephi$.

We may wonder at this stage how such a system compares with a real star and
especially its baroclinicity. In a star, the isothermal and isentropic
surfaces are more spherical than the equipotentials or isobars
\cite[e.g.][]{B82}; since the stellar envelope is stably stratified,
entropy increases outwards which implies that, on an equipotential,
entropy also increases from pole to equator. In our Boussinesq model,
entropy (actually potential temperature) is represented by temperature;
our equipotentials being oblate ellipsoids and the temperature gradient
being positive outwards, we see that there too, entropy increases from pole to
equator of an equipotential surface. We may also note that as shown by
\cite{zahn74}, the baroclinic torque is proportional to the latitudinal
gradient of entropy on an equipotential.

\subsection{Scaled equations}

Gathering vorticity, energy and continuity equations, we need to solve
the following system:

\greq
\na\times\lp 2\vO\wedge\vv + \vv\cdot\na\vv + \alpha\delta
T(\vg+\Omega^2s\es) - \nu\Delta\vv \rp = -\eps N^2(r)\sth\cth\ephi\\
\vv\cdot\na T_{\rm eq}+\vv\cdot\na \delta T = \kappa\Delta\delta T\\
\Div\vv=0
\egreq
since this is a forced problem, we wish to have a forcing of order unity
as well as the solution. These considerations lead us to the following
scaling of the velocity field and temperature perturbations:

\[ \vv = \frac{\Omega{\cal N}^2R^2}{2g}\vu, \qquad \delta T = \eps T_*
\theta \quad {\rm with} \quad {\cal N}^2 = \frac{\alpha T_*g}{R}\]
where $\cal N$ is the scale of the \BVF. This scaling takes into account
the fact that the baroclinic flow and the associated temperature
perturbations vanish when either rotation or \BVF\ vanishes. We thus find
the system of dimensionless dependent variables:

\greq
\na\times\lp \ez\wedge\vu + \RO\,\vu\cdot\na\vu -
(r\er-\eps s\es) - E\Delta\vu \rp = -n_T^2(r)\sth\cth\ephi\\
(n_T^2/r)u_r + \eps\vu\cdot\na\theta = \tE_T\Delta \theta\\
\Div\vu=0
\egreqn{first_full_eq}
where we introduced the numbers:

\[E = \frac{\nu}{2\Omega R^2}, \qquad \tE_T=\frac{\kappa}{2\Omega
R^2}\lp\frac{2\Omega}{\cal N}\rp^2,
\qquad \RO = \frac{{\cal N}^2R}{4g}\]
$E$ is the Ekman number which measures the viscosity, $ \tE_T$ measures
heat diffusion and Ro is the Rossby number. In addition to these numbers
we will need the Froude number, Fr, the Prandtl number $\PR$ and the
$\lambda$-parameter introduced by \cite{garaud02}; these are respectively:

\[ \FR = \frac{V}{{\cal N}R} = \frac{\Omega {\cal N} R}{2g}, \qquad
\PR=\frac{\nu}{\kappa}, \qquad \lambda = \frac{E}{\tE_T} =
\PR\frac{{\cal N}^2}{4\Omega^2}\]
Finally, $n^2_T(r)$ is the scaled \BVF.

\subsection{Boundary conditions}

This systems needs to be completed by boundary conditions. We shall
assume the regularity of the solutions at the sphere's centre and impose
stress-free boundary conditions on the velocity field at the outer surface.
Thus doing, the velocity field is determined up to a solid rotation;
if $\vu$ is a solution of \eq{first_full_eq} then $\vu+A\ez\times\vr$
is also a solution ($A$ is an arbitrary constant). For actual stars,
such a degeneracy is lifted by initial conditions and conservation of
angular momentum. Here we lifted it
by imposing that the solution $\vu$ of \eq{first_full_eq} has no total
angular momentum, i.e.  that

\[\intvol su_\varphi dV =0 \]
so that the total angular momentum of the star is in the background
solid rotation measured by $\Omega$ (which therefore appears as the mean
rotation rate).

We further complete \eq{first_full_eq} by also imposing zero temperature
fluctuation on the outer surface.

\subsection{Discussion}

As can be seen, the problem is controlled by a large number of
parameters namely

\begin{itemize}
\item the ratio of the centrifugal acceleration $\eps$ to surface
gravity which is also the ellipticity of equipotentials,
\item the Rossby number Ro,
\item the diffusion coefficients, $E,\tE_T$,
\item the profile of \BVF\ $n_T^2(r)$.
\end{itemize}

Moreover, two other parameters will be necessary to described molecular
weight gradients, namely the \BVF\ profile $n_\mu^2(r)$, and the related diffusion
coefficient, while another one will characterize the viscosity jump at
the core-envelope interface. We therefore need some guidance in this
large parameter space.

\begin{figure*}
\begin{minipage}[c]{0.5\textwidth}  
\centerline{\includegraphics[width=8cm,angle=0]{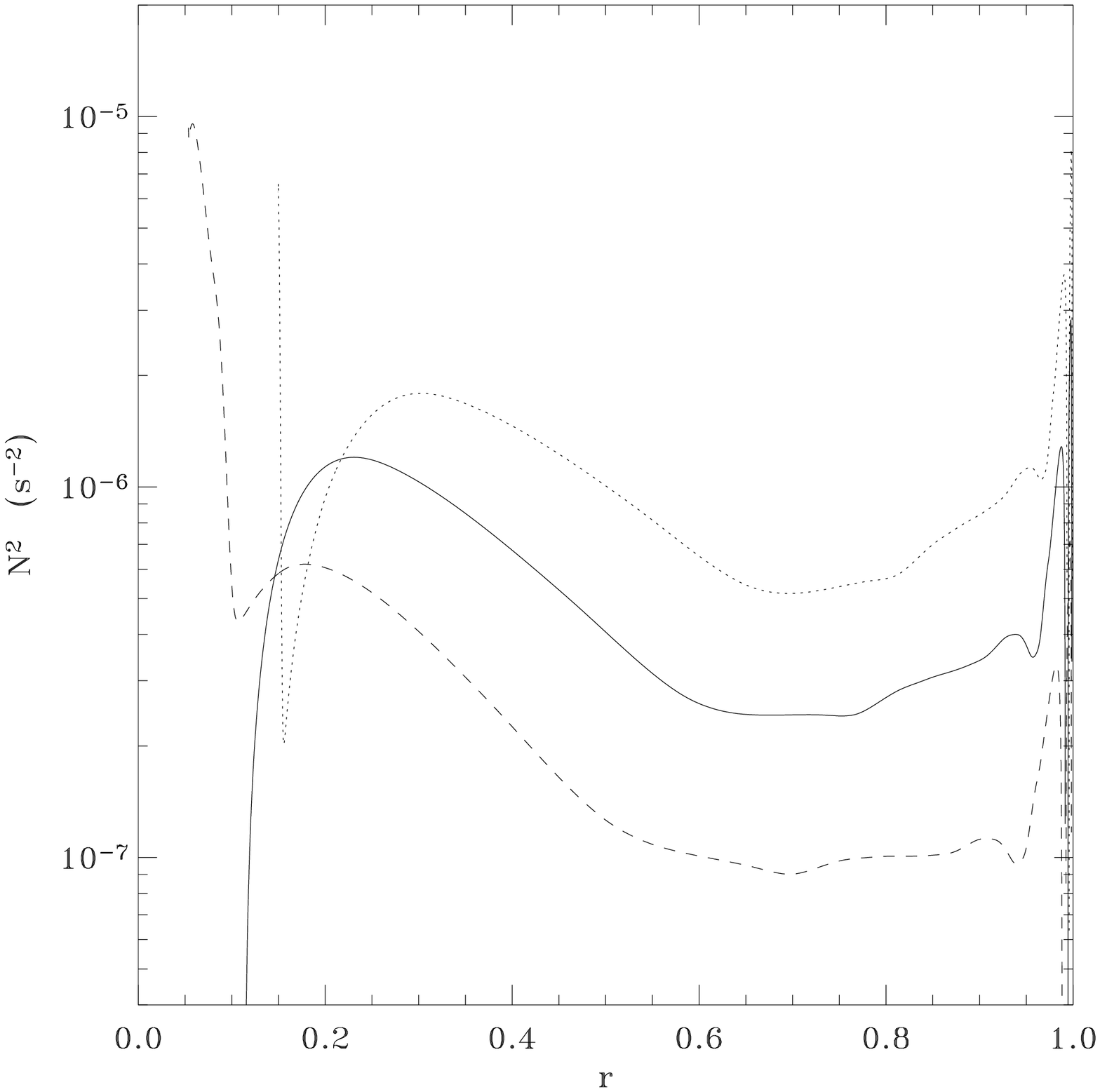}}
\end{minipage}\begin{minipage}[c]{0.5\textwidth}
\includegraphics[width=8cm,angle=0]{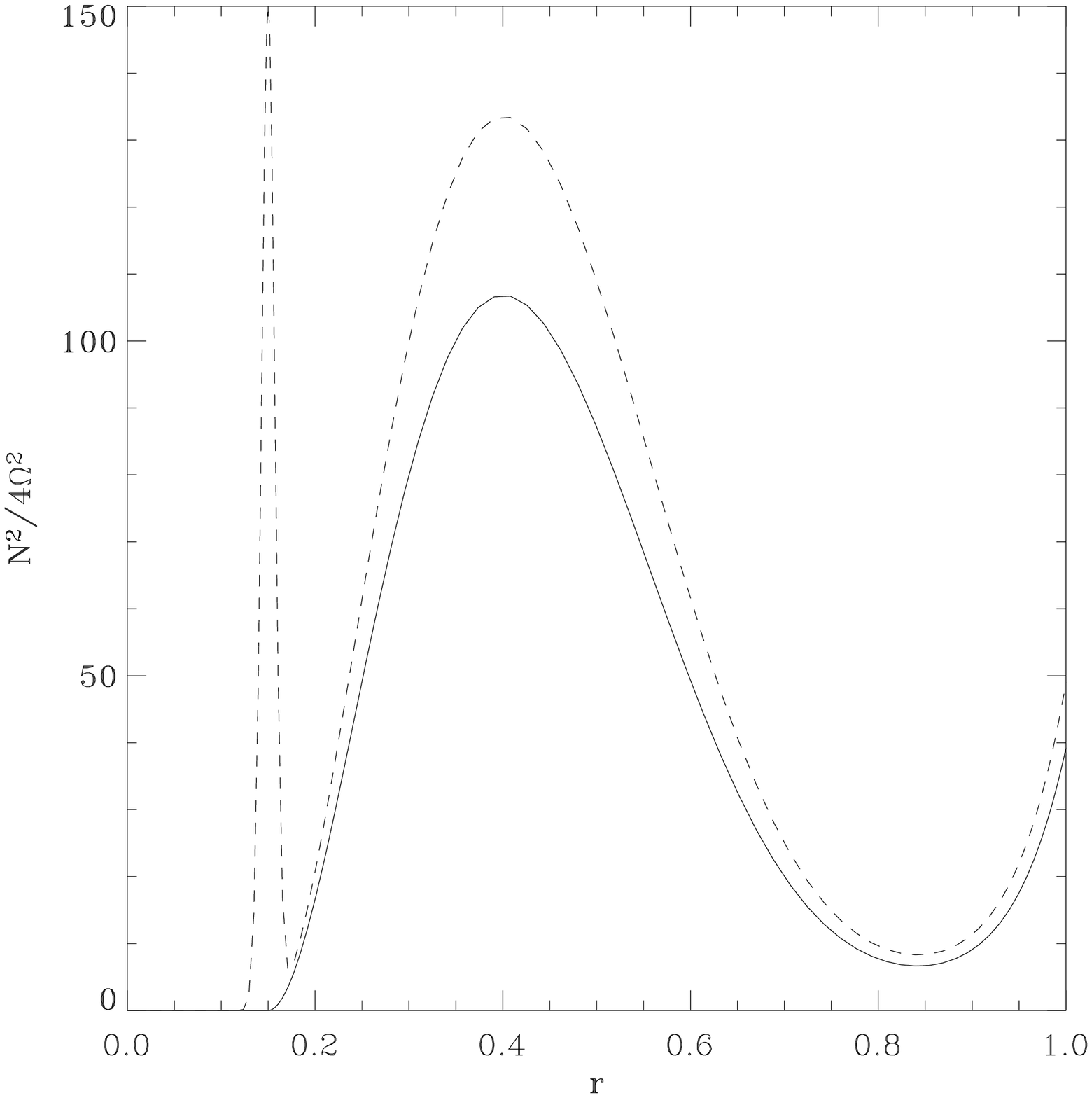}
\end{minipage}
\caption[]{(a) Profiles of \BV\ frequency issued from the 1D evolution
code CESAM for a 3 M$_\odot$ star at 0 (solid), 20 Myrs (dotted line),
284~Myrs (dashed line); these models include microscopic diffusion of
\cite{MP93}. (b) Adopted profiles for a star with (dashed line) and without
(solid line) $\mu$-barrier.}
\label{probv_star}
\end{figure*}

For this purpose and in order that this model enlighten us on real systems,
we shall consider the case of a 3 M$_\odot$ star with a one day rotation
period. Thus we will use a radius $R=2\, 10^9$m, a typical \BVF\ of
${\cal N}=10^{-3}$ Hz and a surface gravity of $g=10^2$ m/s$^2$. Besides,
the profile of the \BV\ frequency needs to reflect the more realistic
models. In fig.~\ref{probv_star}a we plot such a profile for a 3M$_\odot$
star at different stage of its evolution on the main sequence. Such
profiles should not be taken at face value especially the contribution
of the $\mu$-gradients since only microscopic diffusion is included in
this model (produced by the code CESAM, see \cite{Morel97}). In fact it
is the aim of the present work to understand the mechanisms by which
elements move in the radiative envelope. Therefore we shall rather consider
the generic situation visualized in Fig.~\ref{probv_star}b.

\begin{figure}
\centerline{\includegraphics[width=8cm,angle=0]{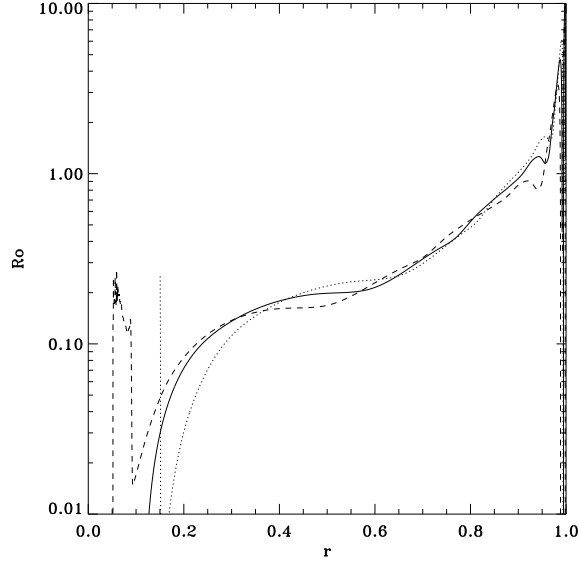}}
\caption[]{(a) Profiles of  the Rossby number $rN^2(r)/4g(r)$ issued
from the 1D evolution code CESAM for a 3M$_\odot$ star at 0 (solid),
20Myrs (dotted), 284Myrs (dashed).}
\label{rossby}
\end{figure}

Because of our choice of scalings, 
amplitude of nonlinear terms in the momentum equation is independent of
rotation and can be appreciated directly from a non-rotating stellar
model. In Fig.~\ref{rossby}, we display the values of the Rossby
number as a function of radius for our typical 3M$_\odot$ star. Indeed,
the non-dimensional function

\[ \RO(r) =  \frac{{\cal N}^2(r)r}{4g(r)}\]
gives an interesting approximation of the amplitude of these non-linear terms.
As can be seen, this number is generally less than unity except near the
surface layers where it reaches $\RO\sim 5$. 

Recalling that $\eps<1$, we see that the other non-linear terms are also
less than unity (in fact very much less than unity as shown below). As
a first step in the investigation, we shall ignore them altogether so as
to be able to examine the properties of the system with linear equations.

Setting $\RO=\eps=0$, we obtain the system:

\greq
\rot\lp\ez\wedge\vu - \theta\vr -E\Delta\vu\rp = -n^2_T\sth\cth\ephi \\
\\
(n_T^2/r)u_r  = \tE_T\Delta \theta\\
\\
\Div\vu=0
\egreqn{sys_lin}

\section{The asymptotic analysis at $E\ll1$}

Before solving the full system \eq{sys_lin}, we shall first discuss
the case of asymptotically small Ekman numbers which are met in stellar
applications. 

\subsection{The inviscid profile}

When viscosity is neglected the equations  \eq{sys_lin} admit a
particular solution called (in geophysics) the thermal wind; in our case
it reads

\greq \vu^0 = \lp s\int \frac{n^2(r)}{r}dr + F(s)\rp\ephi,\\
\\
 \theta=0
\egreqn{vt0}
where $s=r\sin\theta$ is the radial cylindrical coordinate and $F(s)$ an
arbitrary function describing a pure geostrophic solution.

This solution deserves some comments. First, as already pointed
out by \cite{Buss81,B82}, the so-called thermal disequilibrium
induced by centrifugal acceleration does not imply meridional
circulation. A slight differential rotation where the baroclinic torque
is balanced by the Coriolis torque gives a steady solution.

However, we also see that such a solution is largely under-determined since
the function $F$ needs to be specified; this degeneracy is lifted by
viscosity. The qualitative importance of $F(s)$ comes from the fact that it
controls the latitudinal differential rotation.

\subsection{The Ekman boundary layer}

The foregoing thermal wind solution usually does not satisfy the viscous
boundary conditions and thus a boundary layer correction needs to be
added. Let us review here its main properties.

If $\vu^0$ is the inviscid, interior solution and $\tu$ the boundary
layer correction such that $\vu^0+\tu$ satisfies both the flow equations
and the boundary conditions, it is well-known
\cite[see][]{Green69,Rieu97} that $\tu$ verifies

\[ \er\wedge\tu+i\tu = \vC(\vu^0)\exp\lp-\zeta\sqrt{i|\cth|}\rp\]
where $\zeta=(1-r)/\sqrt{E}$ is the radially stretched coordinate and $\vC$ a
complex vector. In the boundary layer the flow is essentially tangential
and we may write the full velocity field:

\[ u_\theta+iu_\varphi = C\exp\lp-\zeta\sqrt{i|\cth|}\rp + iu_\varphi^0\]
where the constant $C$ is such that the horizontal stress is zero; namely

\[ \dr{}\lp\frac{u_\theta+iu_\varphi}{r}\rp=0\]
which yields

\[ C\equiv C(\theta)= (1+i)\sqrt{\frac{E}{2}}\Gamma(\theta)\]
with 

\begin{equation}
\Gamma(\theta) = \frac{F(\sth)-\sth F'(\sth)-n^2(1)\sth}{\sqrt{|\cth|}}
\end{equation}
where the prime indicates derivatives. From these results we find the
meridian velocity  in the boundary layer:

\begin{equation}
u_\theta = \sqrt{\frac{E}{2}}\Gamma(\theta)(\cos\xi+\sin\xi)e^{-\xi}
\label{uth}
\end{equation}
with $\xi=\zeta \sqrt{|\cth|/2}$.

Such a flow however does not verify mass conservation. This is taken care
of by the so-called Ekman pumping which yields the radial component:

\begin{equation}
\tilde{u}_r=EU(\sth)\cos\xi e^{-\xi}
\end{equation}
with

\[U(s) = n^2(1)(1+q(s))+sF''(s)+q(s)(F'-F/s), \qquad {\rm with}\qquad q(s) =
1+\frac{s^2}{2(1-s^2)} \]

These expressions show us that near the surface there exists a
latitudinal flow ${\cal O}(\!\sqrt{E})$ which induces an \od{E} circulation
in the bulk of the fluid; in turn, this circulation controls the
geostrophic flow $F(s)\ephi$ and thus removes the degeneracy of the
thermal wind solution. The reader may have noticed that the boundary
layer solution is singular at equator ($\theta=\pi/2$); this is the
classical equatorial singularity of Ekman layers whose thickness changes
to \od{E^{2/5}} in a region \od{E^{1/5}} around the equator \cite[see
also][]{Green69}.

\subsection{The thermal wind}

Since the differential rotation is a major feature of  the baroclinic
flow, it is useful to push further its analysis and try to express $F(s)$
as a function of the \BV\ frequency profile.

For this purpose, we need to express the meridian circulation as a
function of $F$. We thus write the $\varphi$-component of the momentum
equation in cylindrical coordinate

\[ u_s = E(\Delta -1/s^2)u^0_\varphi \]
where $\Delta$ is the laplacian operator; this equation expresses the
local balance between advection and diffusion of angular momentum. It
yields

\beq u_s = E(F''+F'/s -F/s^2 + sC(r)), \qquad {\rm with}\qquad C(r) =
\lp\frac{n^2}{r}\rp'+ \frac{4n^2}{r^2}\eeqn{us}
Mass conservation gives the z-component of the flow:

\beq u_z = -\int^z_0 \frac{1}{s}\ds{su_s}dz = -EL_3(F)z- E\int^z_0
\frac{1}{s}\ds{s^2C(r)} dz'
\eeqn{uz}
with $L_3(F) = \lc(sF')''-(F/s)'\rc\big/s$ while in the integral $r^2=z'^2+s^2$.

Now, we put all the pieces together by demanding that $u_r=0$ is satisfied
at order \od{E} on the outer boundary. Hence, setting $r=1$, we have

\[ su_s + zu_z+\tilde{u}_r = 0\]
which gives the following differential equation for $F$:

\beq sF''+F'-F/s+s^2C(1)-\zeta(s)^2L_3(F) - \zeta(s)\int_0^{\zeta(s)}
\lp 2C(r)+s^2\frac{C'(r)}{r}\rp dz + n^2(1)(1+q(s))+sF''(s)+q(s)(F'-F/s) =
0\eeqn{eqgeo}
with $\zeta(s)=\sqrt{1-s^2}$.  This complicated equation is of the form
${\cal L}(F) = b$, namely a forced third order differential equation. We
shall not try to find the general solution of it, but will examine the
polynomial solution $F=As+a s^3 + bs^5$, formally valid at $s\ll1$, with
the hope that since $s\leq1$, it will give a rather good idea of the flow.

We first note that, as expected, the solution is invariant to the addition of a
solid rotation; thus $A$ is arbitrary. The first interesting term is $as^3$.
Substitution into \eq{eqgeo} give the following result:

\[ a=-\demi\int_0^1\frac{n^2(r)}{r^2}dr, \qquad b =
\frac{1}{48}\lp\frac{n^2}{r}\rp'(1) + \frac{19}{192}n^2(1) -
\frac{1}{24}\int_0^1\la\frac{n^2}{r^2}+\frac{1}{3r}\lc\lp\frac{n^2}{r}\rp'' +
4\lp\frac{n^2}{r^2}\rp'\rc\ra dr \]
and thus we expect the following differential rotation close to the
z-axis

\beq \delta\Omega =  \int \frac{n^2(r)}{r}dr -
\frac{s^2}{2}\int_0^1\frac{n^2(r)}{r^2}dr + b s^4
\eeqn{rotdiff}
which is now an explicit function of the \BV\ frequency profile.

\begin{figure}[t]
\centerline{\includegraphics[width=9cm,angle=0]{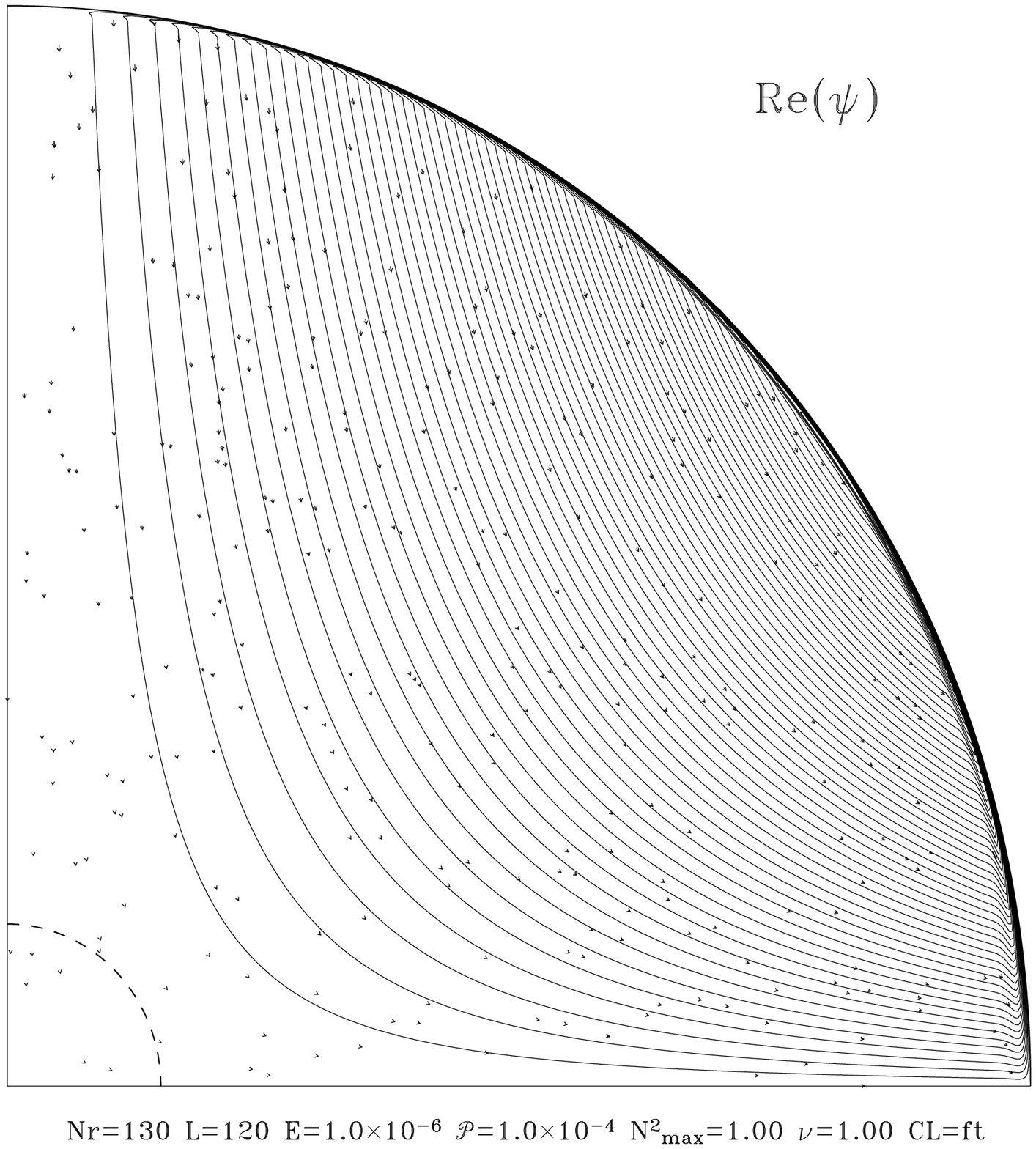}
\includegraphics[width=9cm,angle=0]{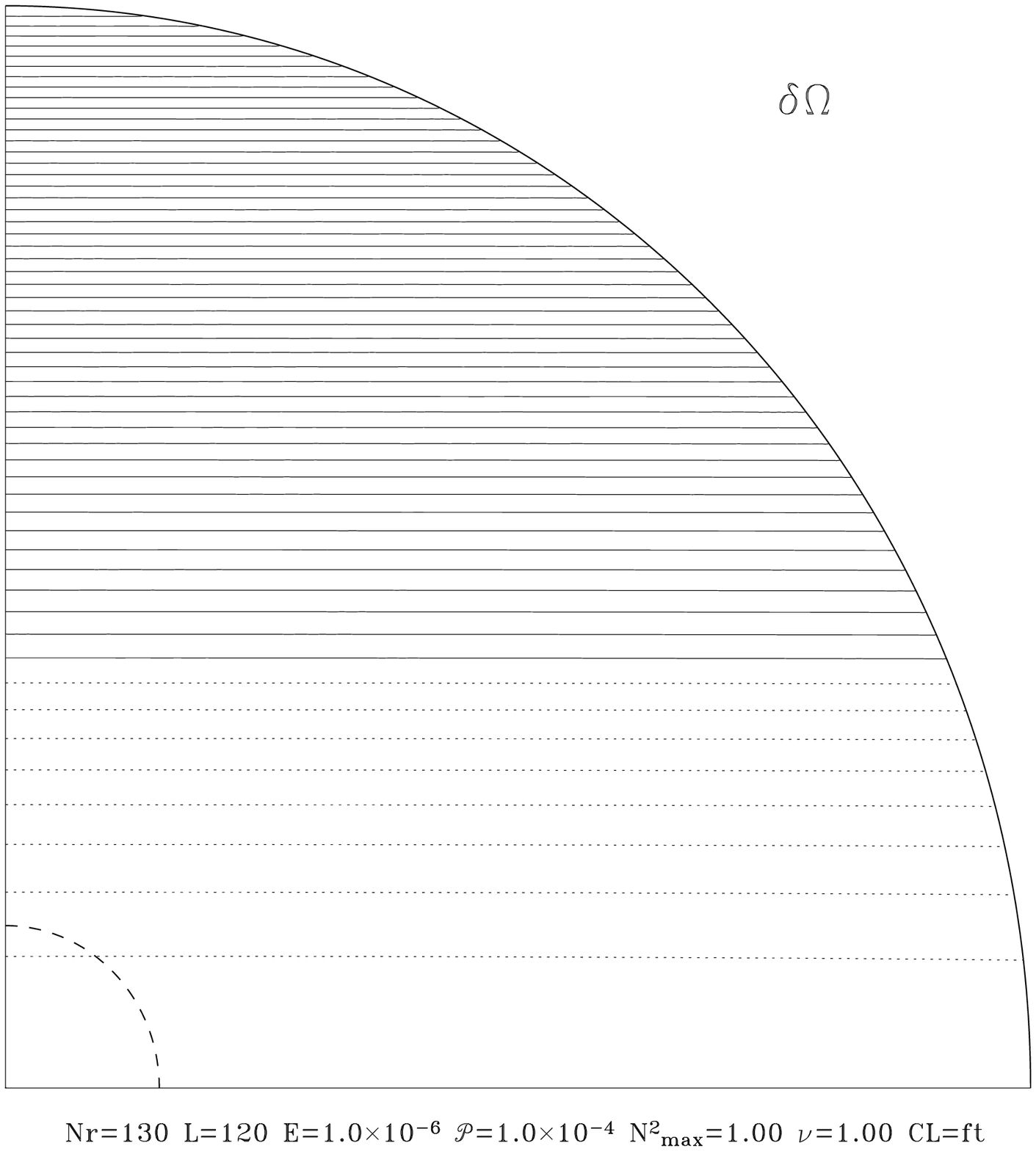}}
\caption[]{Differential rotation (right) and meridional circulation
(left) for a \BVF\ profile $n(r)=r$.}
\label{rot_diff}
\end{figure}

\subsection{The role of stratification}

In the foregoing solution we set $\theta$ to zero and therefore did not
bother about stratification. In order to understand its role, we need to
evaluate the amplitude of $\theta$, i.e. the scaled temperature
fluctuation.

Equation (\ref{sys_lin}b) shows that since $u_r$ (i.e. meridian
circulation)
is \od{E}, then $\theta$ is \od{E/\tE_T} that is of the order of
$\lambda={\cal P}N^2$.  Thus, in the limit of
small $\lambda$'s, our solution is certainly valid. Stellar fluids are
characterized by small Prandtl numbers and usually in rapidly rotating
stars one considers that $\lambda\ll1$ which makes our neglect of
buoyancy comfortable.

However, as was pointed out long ago by \cite{SZ70} it is most likely
that some turbulence is driven by the shear of the baroclinic flow.
The viscosity, changed to a turbulent one, is then increased by some
factor which therefore also increases the Prandtl number. From the
discussion of \cite{zahn93}, it turns out that this turbulent Prandtl
number may reach values close to unity; hence, even in a rapidly rotating
star, the $\lambda$-parameter may be larger than unity (typically
up to $10^2$).

The consequence of such a situation is that a thermal boundary layer
appears. Its scale is of course $1/\sqrt{\lambda}$ and we now assume
that $\lambda\gg1$. Outside this layer, quantities vary on an \od{1}
scale; noting that $u_s\sim u_r$, using the $\varphi$-component of the
momentum equation together with the heat equation, it follows that

\[ \theta\sim \lambda u_\varphi\]
Now, the $\varphi$-component of the vorticity equation (\ref{sys_lin}a)
shows that in this region

\[ u_\varphi\sim \lambda^{-1}, \quad u_r \sim E\lambda^{-1}, \quad {\rm
and} \quad \theta\sim 1\]
Roughly speaking, outside the thermal layer the temperature fluctuation
is of order unity and the velocity perturbation is reduced by a factor
\od{\lambda}, while in the thermal layer, radial gradients are increased
by a factor $\sqrt{\lambda}$ which gives:

\[ u_\varphi\sim \lambda^{-1}, \quad u_r \sim E, \quad {\rm
and} \quad \theta\sim 1\]
Hence, all the dynamics described above for $\PR\ll1$ becomes squeezed
inside the thermal layer with a differential rotation reduced by a
factor $\lambda$. This is in fact very similar to the solar tachocline
where turbulent diffusion inhibits the diffusion of momentum with the
help of stratification.

\subsection{Discussion}

The foregoing solution potentially shows very interesting properties of the
baroclinic flow; however, we may wonder if it is reliable mathematically
considering the crudeness of the solution of \eq{eqgeo}.

The accuracy of the $as^3$ solution of \eq{eqgeo} may only be
estimated by direct comparison to numerical solutions. Anticipating
on the next section we solve numerically \eq{sys_lin} in the limit of
small Prandtl number. We used a very simple \BV\ frequency profile,
namely $n^2=r^2$, which is actually quite standard in the literature
\cite[see][]{chandra61}. With such a profile the estimate of the different
components of the flow are quite easy to evaluate ($a=-\demi$ and
$b=\frac{15}{192}$); as for the differential equation we find

\[ \delta\Omega = \demi z^2\]
neglecting $b$; this solution compares extremely well with the numerical
solution (see Fig.~\ref{rot_diff}) as soon as ${\cal P} \leq 0.1$. It
shows that the second term $bs^5$ does not much influence the shape of
the solution. From the expression of $b$ we see that this is true
because, in this case, $n^2$ is a smooth function of $r$. In stellar
situations, the coefficient $(n^2/r)'(1)$ might be very large if sharp
gradients develop in the upper layers of the star.

Reassured by these computations, equation \eq{rotdiff} therefore shows a
very interesting result: {\em if the \BV\ frequency profile of a star is
sufficiently smooth, the latitudinal differential rotation in a stably
stratified envelope driven by the sole baroclinicity is a fast rotating
pole and a slower rotating equator}. Moreover, the $a$-coefficient being
an integral over the star, it is not sensitive to the detailed shape of
the \BV\ profile and therefore is a rather robust quantity.

Conversely, in a convecting envelope, with viscous-like Reynolds
stresses, baroclinic torques induce a fast rotating equator and slow
poles, like the sun actually. Of course, in the solar case, this may
just be a coincidence since baroclinicity in the convective zone of the
sun hardly comes from centrifugal effect but rather from the convection
itself \cite[see][]{BT02}. Nevertheless, in rapidly rotating (i.e. young) solar
type stars, our result shows that baroclinicity drives a solar-like
differential rotation.

Finally, the solution also gives the form of the meridian circulation through
\eq{us} and \eq{uz}. For instance the number of cells can be retrieved from the
expression of $u_s$. At equator, $s=r$ and $u_s$ reads

\[ u_s=E\lp\frac{1}{r^3}\dnr{(r^3n^2)}-4r\int^1_0\frac{n^2}{r^2}dr\rp \]
The number of zeros of this function, plus one, gives the number of cells
according to the \BV\ frequency profile. Besides, the expression of
$u_\theta$ in the Ekman layer \eq{uth} shows that at the very surface,
$u_\theta \sim -n^2(1)s$, which means that the flow is poleward. If
the \BV\ frequency profile is smooth enough this motion extends to
low latitudes. However, because of the $\sin\xi+\cos\xi$
dependence, the boundary layer flow reverses just below the surface at
$r=1-\frac{3\pi}{4}\sqrt{\frac{2E}{|\cth|}}$.

From the amplitude of the meridian flow, one can estimate the circulation
time scale which is:

\[ T_{\rm circ} = \frac{2\Omega}{{\cal N}Fr} T_{\rm visc} = \frac{T_{\rm
visc}}{\RO}\]
Since $\RO\sim1$, we see that this time scale is of the order of the
viscous diffusion time scale $T_{\rm visc}$ which is very large if only
microscopic viscosity is diffusing momentum.

\section{The general case}

We now turn to a more complex model with which we wish to analyse the
effects of a convective core and of $\mu$-gradients. The inclusion of
$\mu$-gradients amounts to the addition of the equation of a concentration
$c$ and the modification of the buoyancy term; the linear system
\eq{sys_lin} needs to be changed into:

\greq
\rot\lp\ez\wedge\vu - (\theta-c)\vr -E\Delta\vu\rp = -n^2\sth\cth\ephi
\\
\\
(n_T^2/r)u_r  = \tE_T\Delta \theta\\ 
\\
-(n_\mu^2/r)u_r = \tE_c\Delta c\\
\\
\Div\vu=0
\egreqn{sys_lin_mu}
where we introduced $n^2=n_T^2+n_\mu^2$ and $\tE_c=(D_c/2\Omega
R^2)\times(2\Omega/{\cal N})^2$ with $D_c$ being the diffusivity of
element ``c". This element is supposed to be heavier than the
surrounding gas.

Since solutions are much more complicated we turn towards numerical
solutions and first describe the numerical technique.

\subsection{The numerical method}

To solve the system \eq{sys_lin_mu}
we first project the variables onto the spherical harmonic base
\cite[see][]{rieu87}

\[\vu=\sum_{\ell=0}^{+\infty}\sum_{m=-\ell}^{+\ell} \ulm\RL+\vlm\SL+\wlm\TL, \qquad 
\theta=\sum_{\ell=0}^{+\infty}\sum_{m=-\ell}^{+\ell} \theta^\ell_m\YL, \qquad
c=\sum_{\ell=0}^{+\infty}\sum_{m=-\ell}^{+\ell} c^\ell_m\YL \]
where

\[\RL=\YL\vec{e}_{r},\qquad \SL=\na\YL,\qquad \TL=\na\times\RL \]
with $\YL$ being the usual normalized spherical harmonic function. Since
the flow is divergenceless, $v^l_m$ may be expressed as:

\[ \vlm = \frac{1}{\llp}\frac{1}{r}\dnr{r^2\ulm}\]
Then, projecting the equations (the vorticity equation needs be
projected only on $\RL$ and $\TL$), we get

\[\left\{ \begin{array}{l} 
E\Delta_\ell\wlm+A(\ell)r^{\ell-1}\drr\biggl(
\frac{u_m^{\ell-1}}{r^{\ell-2}}\biggr) 
 +A(\ell+1)r^{-\ell-2}\drr\biggl( r^{\ell+3}u_m^{\ell+1}\biggr) =0  \\
\\ 
E\Delta_\ell\Delta_\ell(r\ulm)=
 B(\ell)r^{\ell-1}\drr
\biggl(\frac{w_m^{\ell-1}}{r^{\ell-1}}\biggr) 
 + B(\ell+1)r^{-\ell-2} \drr\biggl( r^{\ell+2}w_m^{\ell+1}\biggr)
 +\llp(\thetalm-\clm) - (n_T^2+n_\mu^2) N_2\delta_{\ell2} \\
\\
\tE_T\Delta_\ell\thetalm  -n_T^2/r\ulm =0 \\
\\
\tE_c\Delta_\ell \clm  +n_\mu^2/r\ulm =0
\end{array} \right. \]
where $N_2=\sqrt{\frac{16\pi}{5}}$ and

\[A(\ell)=\frac{1}{\ell^2}\sqrt{\frac{\ell^2-m^2}{(2\ell-1)(2\ell+1)}}, \qquad
B(\ell) = (\ell^2-1)\sqrt{\frac{\ell^2-m^2}{(2\ell-1)(2\ell+1)}} \]

This coupled system of differential equation needs to be completed by
boundary conditions at the surface of the sphere (at the centre we only
demand the regularity of the solutions). Concerning the velocity, we
impose stress-free conditions, thus

\[ \ulm=0,\qquad \ddnr{r\ulm}(1) =0, \qquad \dnr{}\lp\frac{\wlm}{r}\rp = 0\]
while on the temperature and concentration we impose respectively no
fluctuation and no flux at $r=1$, namely

\[ \thetalm =0, \qquad \dnr{\clm} = 0\]

Further, we shall need interface conditions at the core-envelope
boundary. These conditions express the continuity of the velocity,
temperature, concentration, fluxes and stress fields. The continuity of
first quantities simply translates as the continuity of

\begin{equation}
\ulm,\quad \dnr{\ulm}, \quad \wlm, \quad \thetalm, \quad \clm
\end{equation}
while the continuity of fluxes and stresses must reflect the changes in
the transport coefficients between the turbulent convective core and the
radiative envelope. Continuity of the three components of stress demands
the continuity of

\begin{eqnarray*}
&& E\lp r\ddnr{\ulm}+2\dnr{\ulm}+ (\llp-2)\ulm/r\rp \\
&& E\lp \dnr{\wlm}-\frac{\wlm}{r}\rp\\
&& \plm-2E\dnr{\ulm} \qquad
{\rm or}\qquad E\lp r^2\dddnr{\ulm}+6r\ddnr{\ulm}+3(2-\llp)\dnr{\ulm}\rp
\end{eqnarray*}
while continuity of heat flux and concentration flux impose the
continuity of 

\[ \tE_T\dnr{\thetalm} \qquad {\rm and} \qquad \tE_c\dnr{\clm} \]

\begin{figure*}
\centerline{\includegraphics[width=9cm,angle=0]{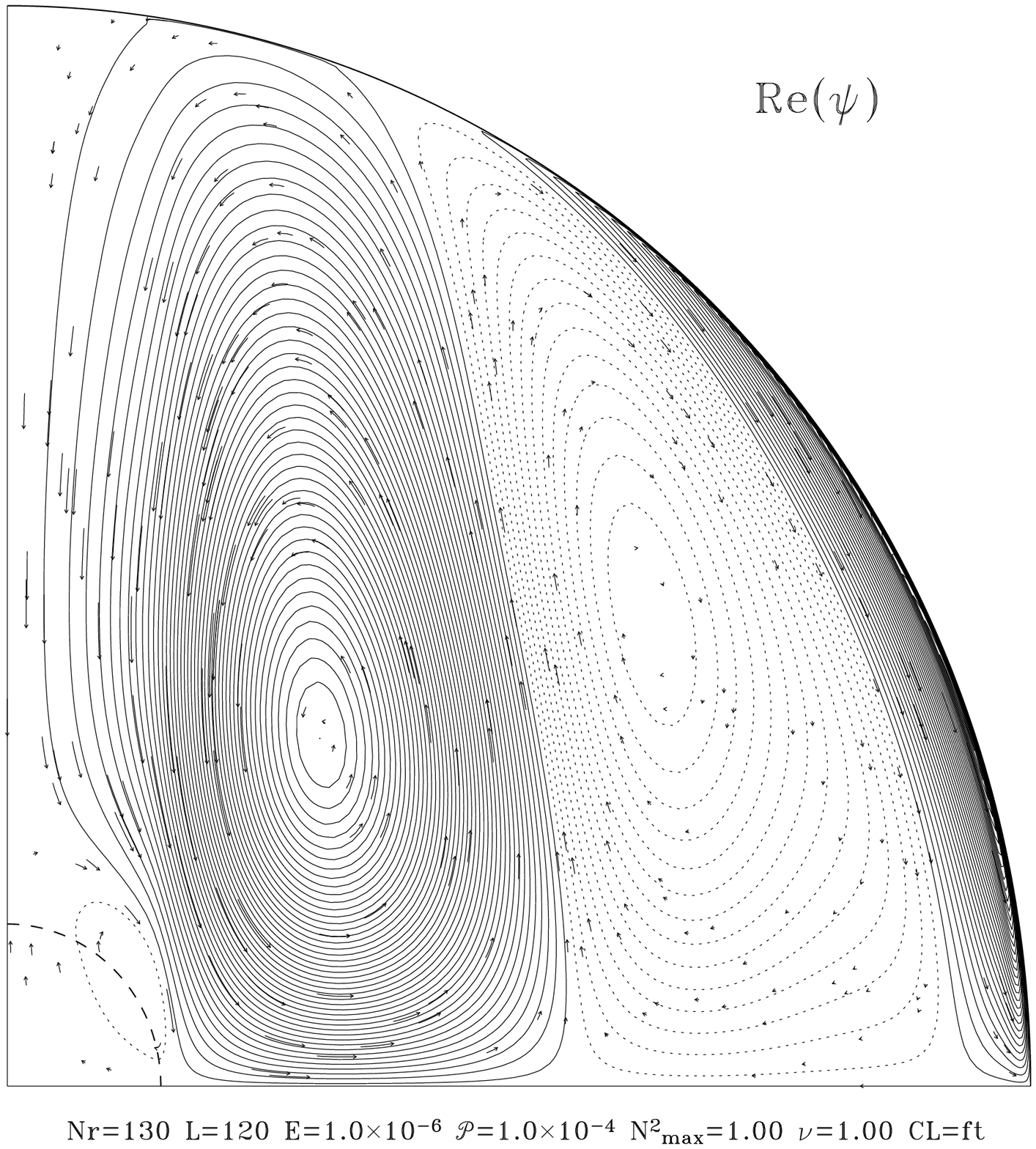}
\includegraphics[width=9cm,angle=0]{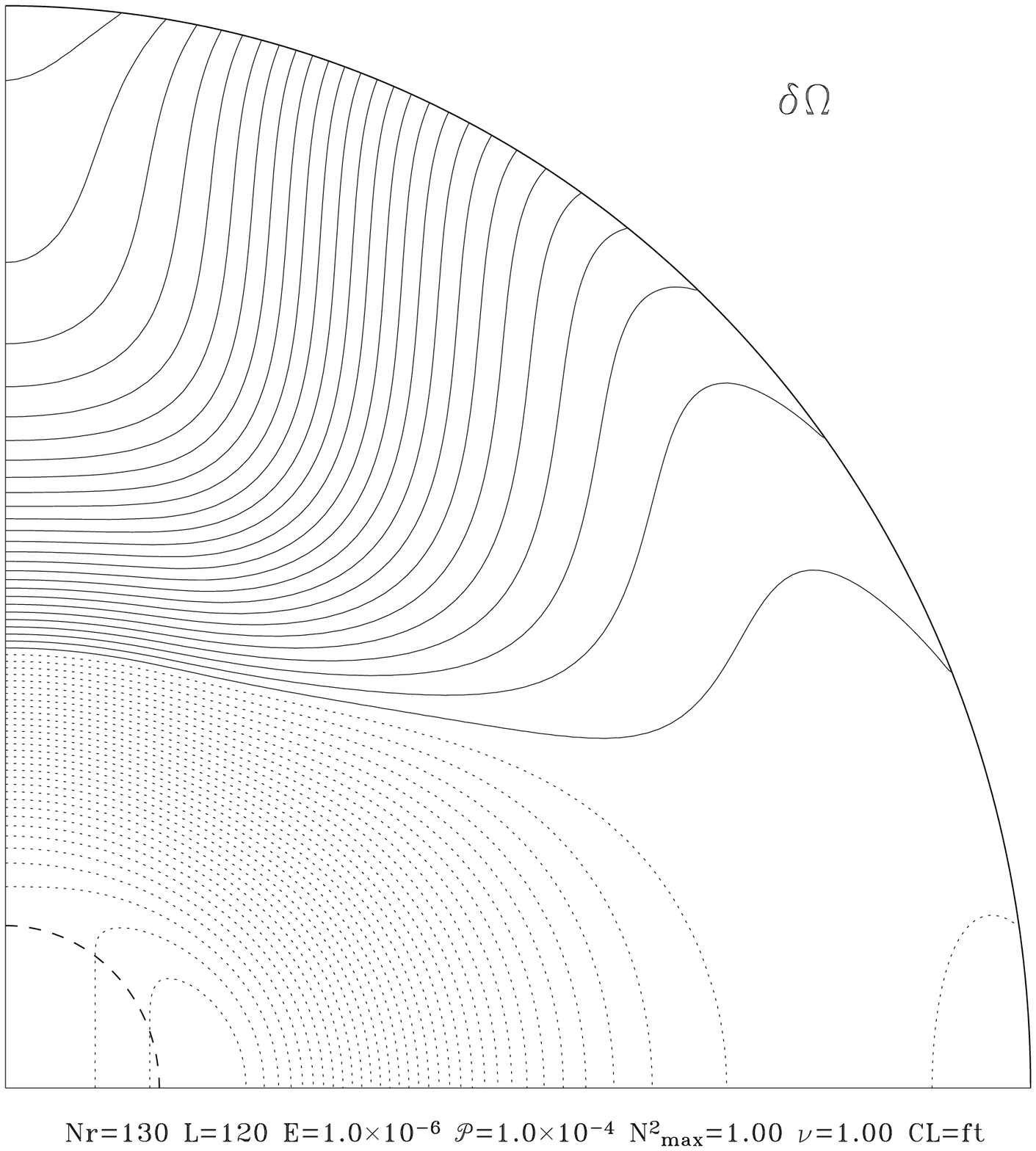}}
\caption[]{Left: the meridional circulation of the baroclinic flow
generated by the \BV\ frequency profile in figure~\ref{probv_star}b
(solid curve) for a fluid with constant viscosity. Right: The associated
differential rotation showing the fast rotating pole and slow equator
(solid lines are for positive contours, dotted lines for negative ones).}
\label{psi1}
\end{figure*}

\subsection{The shape of the baroclinic flow}

We first investigate the baroclinic flow which is represented by both
the differential rotation and the associated meridian circulation. We
focus our attention on the influence of the main uncertainties of the
problem, namely the sensitivity to viscous transport and the height of
the $\mu$-barrier.

\subsubsection{With no jump in viscosity}

\begin{figure}
\hfil
\begin{minipage}[t]{.45\linewidth}
\centerline{\includegraphics[width=9cm,angle=0]{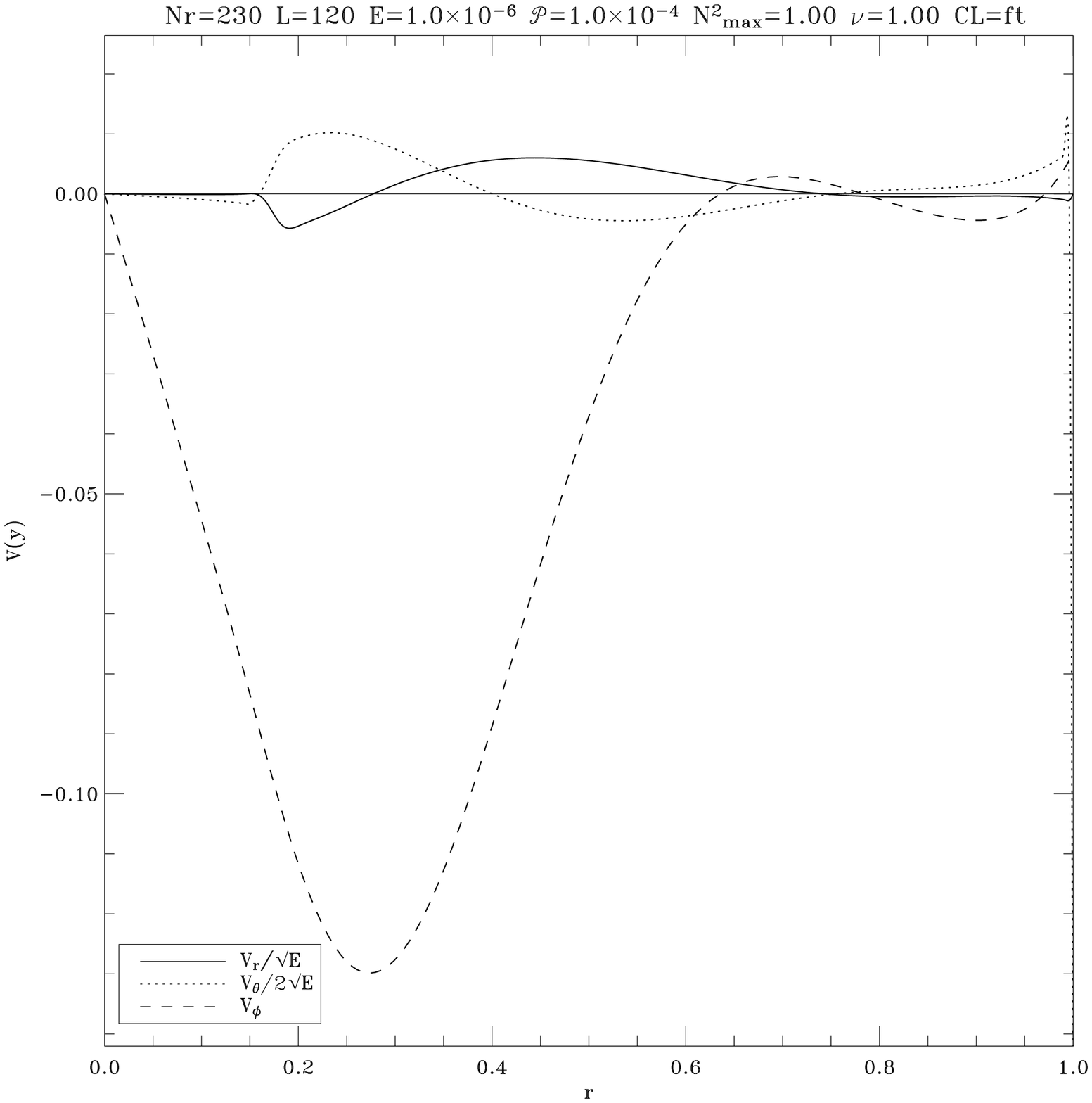}}
\caption[]{Radial profile at $\theta=1$rd of the velocity components
of the baroclinic flow. Note that the $v_r$ and $v_\theta$ have been
multiplied by a factor \od{E^{-1/2}}$\gg1$.}
\label{profil_int}
\end{minipage}
\hfil
\begin{minipage}[t]{.45\linewidth}
\centerline{\includegraphics[width=9cm,angle=0]{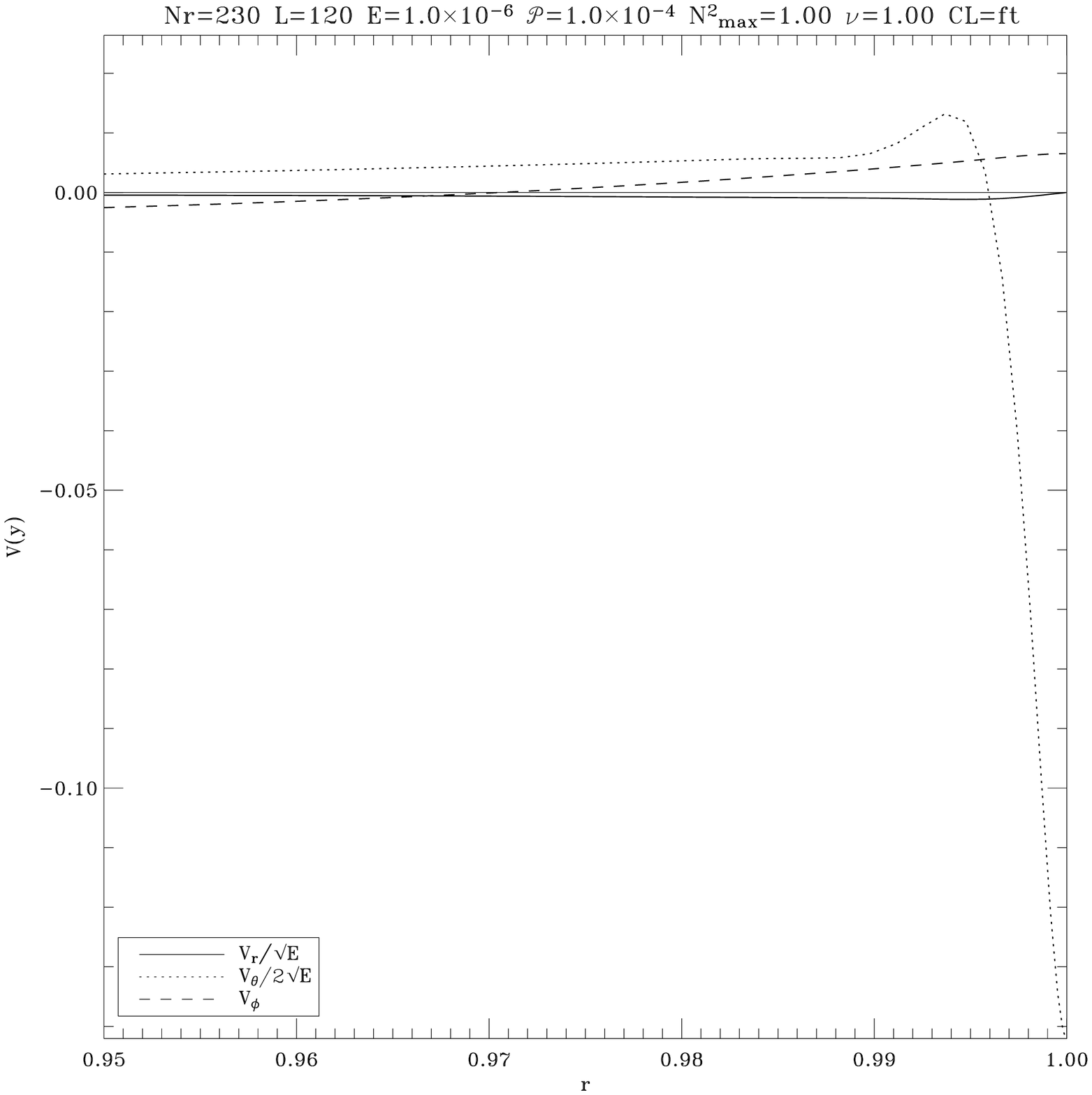}}
\caption[]{Same as Fig~\ref{profil_int}, but with emphasis on the Ekman layer.}
\label{profil}
\end{minipage}
\end{figure}

We first consider a simple configuration where we set a \BV\
frequency profile with no $\mu$-barrier, no viscosity jump at
core-envelope interface. We thus
find a plain baroclinic flow and can check the agreement with our
foregoing theoretical results. The meridian circulation is shown in
Fig.~\ref{psi1} together with the associated differential rotation. The
plot of the angular velocity show the typical shape of the thermal wind
solution given by \eq{vt0}. This shape remains identical if the viscosity
is decreased (i.e. with lower Ekman or Prandlt numbers). This flow is
essentially azimuthal with faster rotating poles as predicted above. The
meridian circulation is much weaker (see Fig.~\ref{profil_int}) being
essentially \od{E}. As expected from the above boundary layer analysis,
near the surface the latitudinal component increases very strongly by a
factor $E^{-1/2}$ (this is essentially an effect of mass conservation). A
detailed view of the velocity profiles in the Ekman layer is given in
Fig.~\ref{profil}.

\begin{figure*}
\centerline{\includegraphics[width=9cm,angle=0]{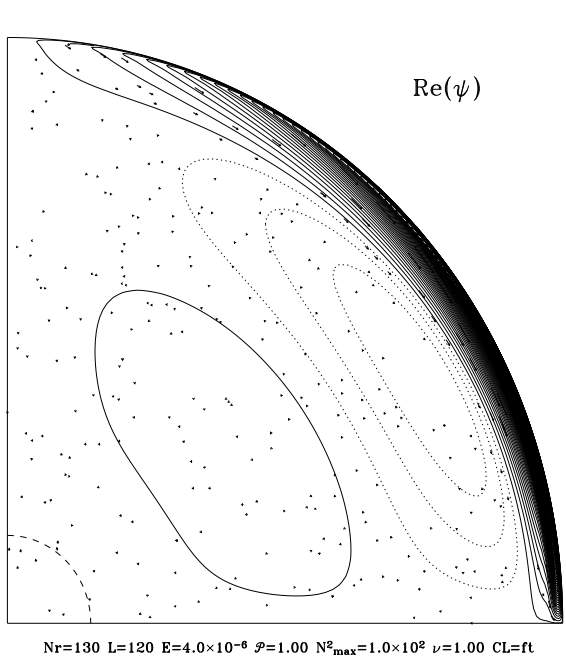}
\includegraphics[width=9cm,angle=0]{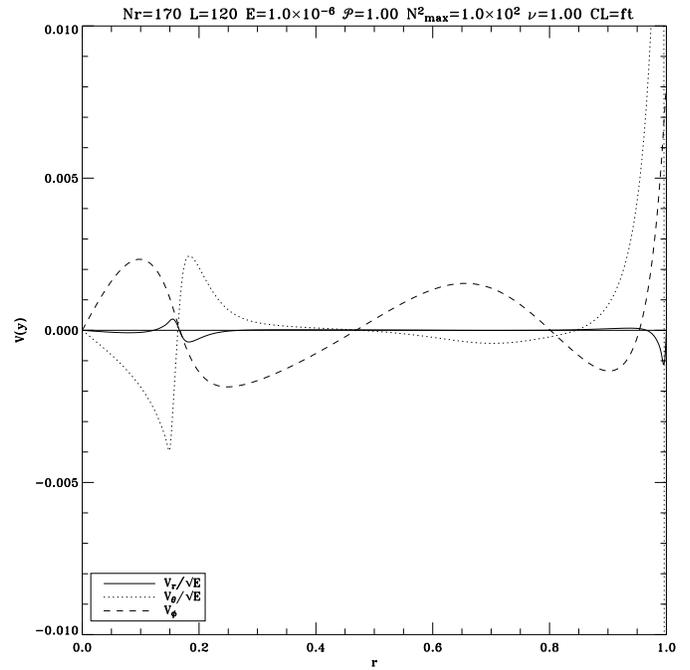}}
\caption[]{Left: Same as Fig.~\ref{psi1}a but with ${\cal P}=1$.
Right: The associated velocity profiles. Note the squeezed circulation in the
thermal layer which width is $\sim0.1$, and the reduction of the azimuthal
velocity in the bulk of the domain.}
\label{psi2}
\end{figure*}

To have further information on the baroclinic flow we had a look to the
case where the Prandtl number is of order unity. In that case the value
of $\lambda$ is important and we set it to 100 according to our typical
rotating star. The meridian circulation along with the radial profiles of
the velocity field are shown in Fig~\ref{psi2}. Obviously, the circulation
has much decreased. This is a direct consequence of the heat equation
(\ref{sys_lin}b) which imposes a reduction of radial velocity when heat
diffusion decreases; as a consequence advection of momentum is less and
therefore diffusion of momentum must reduce which is obtained by reducing
the differential rotation as shown in Fig~\ref{psi2}b.

\subsubsection{With a viscosity jump at the core-envelope boundary}

In our quest of more realistic models, we now consider the effect of
the viscosity jump at the core-envelope boundary which is brought about
by turbulent convection in the core.  The convective core is thus
considered as a much more viscous fluid with negligible stratification
(\BV\ frequency is set to zero).

\begin{figure}
\begin{minipage}[t]{.49\linewidth} 
\centerline{
\includegraphics[width=8cm,angle=0]{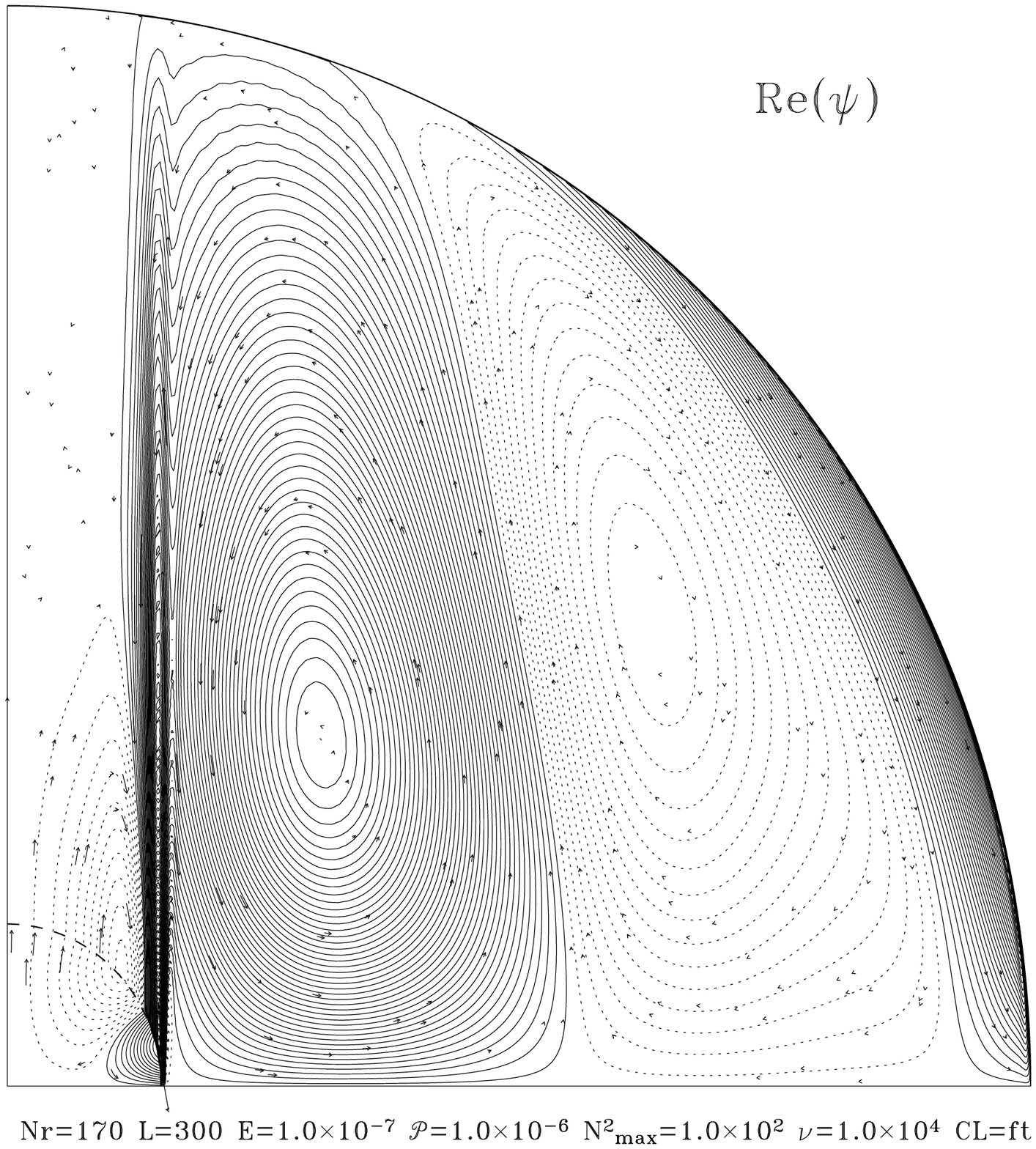}}
\caption[]{Same as figure~\ref{psi1} but the core viscosity is 10$^4$
higher than that of the envelope.}
\label{nomu}
\end{minipage}
\hfil
\begin{minipage}[t]{.49\linewidth} 
\centerline{
\includegraphics[width=8cm,angle=0]{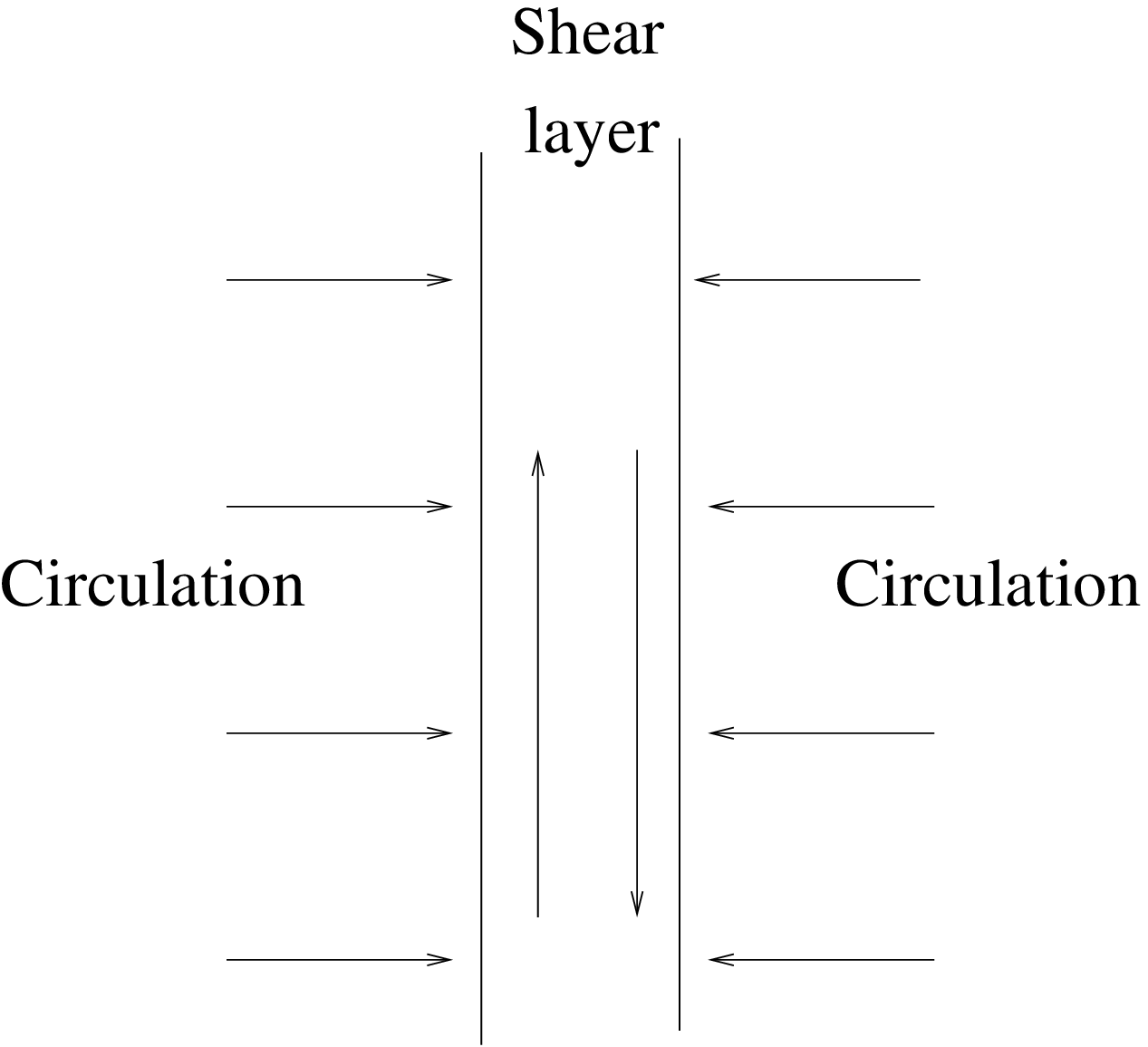}}
\caption[]{Schematic view of the Stewartson layer.}
\label{schem}
\end{minipage}
\end{figure}

The interesting result is that the meridian circulation is strongly
modified as shown in Fig~\ref{nomu}.  This figure shows that the jump
in the mean mechanical properties of the fluids generates a shear layer
parallel to the axis of rotation. Such a layer is a Stewartson layer
well-known in the dynamics of rotating fluids. The dynamics of these
layers is controlled by a delicate balance between viscous stress,
pressure gradient and the Coriolis acceleration. As shown long ago by
\cite{stewar66}, such layers are in fact nested layers whose width scales
with $\nu^{1/3}, \nu^{1/4}$ or $\nu^{2/7}$. However, if the properties of
such layers are well known in the simple case of incompressible fluids,
they remain unexplored when the fluid is stably stratified like here.

\begin{figure}
\centerline{\includegraphics[width=8cm,angle=0]{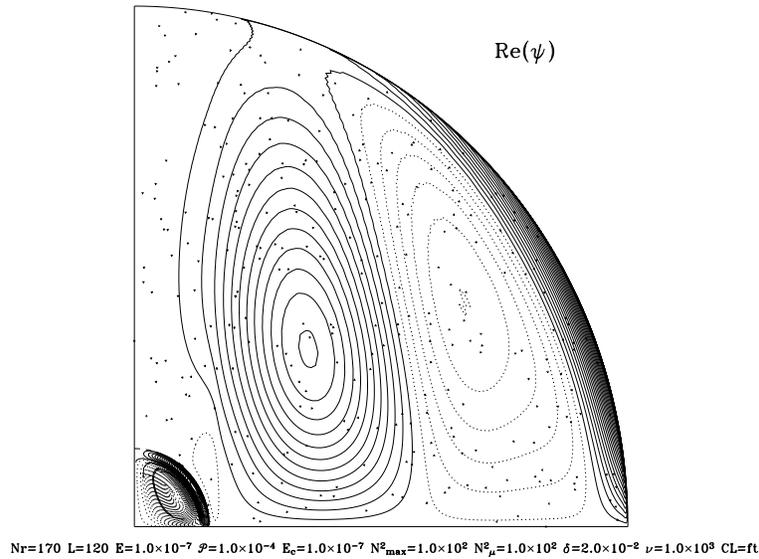}}
\caption[]{Same as Fig.~\ref{nomu} but with a $\mu$-barrier where
$N^2_\mu=N^2_T=100$.  Note the disappearance of the Stewartson layer.}
\label{mu}
\end{figure}

The interesting point of this feature is that the scale of the meridian
flow is much reduced in one direction; the scale controlling the shear
layer pumping is \od{E^{1/4}} if we follow Stewartson theory. This means
that the circulation time is reduced by a factor \od{E^{1/4}}.
Indeed, mass conservation implies that

\[ \frac{\partial V_\perp}{\partial x_\perp} = - \frac{\partial
V_\para}{\partial x_\para}\]
where $V_\perp$ is the pumping induced by the shear layer and which
should be identified to the meridian flow (see figure~\ref{schem}). Since
$\partial/\partial x_\perp \sim E^{-1/4}$ and $\partial/\partial x_\para
\sim 1$, it turns out that $V_\para \sim E^{-1/4} V_\perp$; hence the
advection time from core to surface is reduced by a factor \od{E^{1/4}}
which is quite small as shown below.

Now we may wonder whether this dynamical feature will resist to the
building of $\mu$-barriers during the evolution of the star. To answer
this question, we computed a model with a $\mu$ gradient as shown
in Fig.~\ref{probv_star}b (dashed line); namely the height of the
$\mu$-barrier has been raised such that $N^2_\mu\sim N^2_T$. As far as elemental
diffusivity is concerned we assumed that it is of the same order of
magnitude as the viscosity, i.e. the Schmidt number $\nu/D_c$ is of
order unity.  Figure~\ref{mu} shows the result: the Stewartson layer
has disappeared.  It is replaced by a meridional flow within the core.

\subsection{Questions of stability}

After the computation of the global baroclinic flow, we may wonder
whether it is stable or not. This question is not an easy one since many
instabilities are possible \cite[see][]{KS82} and the investigation
would require a dedicated paper.  However, we can discuss qualitatively
the question.

A first class of instabilities are the barotropic ones; they do not take
advantage of the baroclinic state and are basically due to the shear.
The first to be considered is the axisymmetric one; it is also known as
the centrifugal instability and is controlled by the angular momentum
profile. The Rayleigh criterion, taking into account the stable
stratification, says that instability sets in when

\[ \ds{L_z}<0 \quad {\rm at}\;\; r=Cst, \quad \ssi \quad
\frac{1}{\cth}\dtheta{L_z} < 0\]
that is when the angular momentum decreases with the cylindrical radius. The
constraint $r=Cst$ avoids the stabilizing effect of the \BVF\ profile.

\begin{figure}
\centerline{\includegraphics[width=9cm,angle=0]{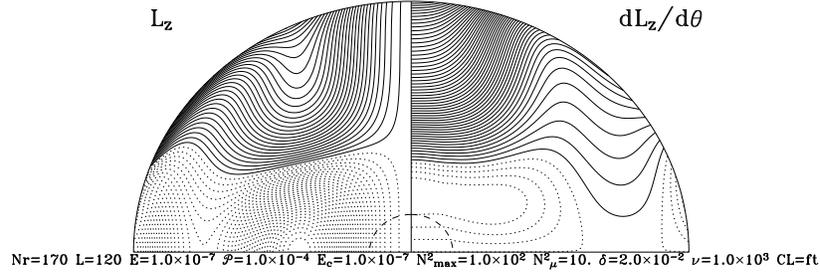}}
\caption[]{Angular momentum distribution and its derivative. The
dotted lines in $\partial L_z/\partial\theta$ show regions where the
barotropic axisymmetric instability may develop (solid lines are for
positive contours, dotted lines for negative ones).}
\label{ang_mom}
\end{figure}

For our system the angular momentum of the flow is

\[ L_z = \Omega s^2 + sV_\varphi = \Omega R^2\lp s^2 + 2\RO\, s\bu_\varphi\rp\]
so that the flow is unstable if

\[ r\cth + \RO\sum_\ell \llp \wlm\YL           < 0\]
This formula shows that the Rossby number controls this instability. As
discussed above, this parameter is in the range [0.1, 10] in our reference
star. We find that for Ro$\infapp3$ the flow is stable and, as shown
in Fig.~\ref{ang_mom}, equatorial regions at destabilized first when Ro
increases. We shall not push further the analysis since at this point
only a detailed analysis would give sensible results. We just conclude
that in view of Fig.~\ref{rossby}, we expect such instability only in
equatorial surface layers.

The other barotropic instability is the non-axisymmetric shear
instability. Basically, in a stably stratified fluid it is controlled
by the Richardson number, e.g.

\[ \RI = \frac{{\cal N}^2}{(dv_\varphi/dr)^2}=\FR^{-2} \]
In our case this number appears to scale with the inverse square of
the Froude number which we found of order unity. We conclude that most
likely the large-scale flow is stable and that a rigorous stability
analysis is needed to give precise answers in stellar conditions. Nevertheless,
following \cite{zahn93}, this flow may still destabilize  scales
small enough that they are insensitive to the stable stratification
because of the large heat diffusion coefficient. The instability of
small scales might thus provide the rotating radiative zone with some
turbulence and hence some turbulent viscosity. \cite{zahn93} shows that
such turbulent viscosity should be within the range

\[ \RE_c\nu \leq \nu_T = \ell^2\dnr{v_\varphi} \leq \frac{\kappa
\RI_c}{{\cal N}^2}\lp\dnr{v_\varphi}\rp^2\]
where Re$_c$ is the critical Reynolds number for shear flow, typically around
1000, and Ri$_c$ the critical Richardson number for stably stratified
shear flow,  around 0.25. These bounds on the turbulent viscosity show
that the viscosity is increased at least by a factor Re$_c$ and at
maximum by a factor $\od{\PR^{-1}}$. Interestingly enough, this raises
the Ekman number from $10^{-18}$ up to $\sim10^{-15},\,10^{-12}$
which is still very small. Moreover, the Prandtl number is
increased by a factor between Re$_c$ and $\PR^{-1}$ which means, in the
latter case, an effective Prandtl number of order unity and a
$\lambda$-number significantly larger than unity. In fact, the
turbulence is likely anisotropic and the values of the effective Prandtl
number depends whether one considers horizontal or vertical diffusion. An
effective Prandtl number of unity is therefore an upper bound.

Besides these classical instabilities (for fluid flows), we have also to
face baroclinic instabilities which are specific to this type of flows
driven by the baroclinicity. As discussed by \cite{SK84}, one should
separate the true baroclinic instability which develops because some
fluid parcel may move in a direction between entropy levels and isobars,
and diffusive instabilities (Goldreich-Schubert-Fricke and ABCD) which
take advantage of double diffusion. These instabilities, specific to
baroclinic flows, are rather weak and one question is whether they
would persist if the diffusivities are increased by some turbulence
generated by the barotropic ones. More work is needed to characterize
them, especially in a global approach.

\section{Discussion}

\subsection{Stellar numbers}

To fix ideas on the foregoing results we now give some numbers  using our
test star. We recall that it is a 3 M$_\odot$ rotating with a period of
one day. On the ZAMS, its radius is 2.6 R$_\odot = 2\, 10^9$m. Typical
kinematic viscosity of the plasma is $\nu=10^{-3}$m$^2$/s yielding an
Ekman number of

\[ E \sim 2\, 10^{-18}\]
really very small. The corresponding viscous time is $T_{\rm visc} =
R^2/\nu \sim 10^{14}$yrs.  The circulation time is typically a viscous
time:

\[ T_{\rm circ} = \frac{R}{v_r} \sim \frac{1}{\FR {\cal N}\,\vu_\varphi}
\sim \frac{T_{\rm visc}}{\RO}\]
since $\RO\sim1$.  As we discussed above, various instabilities may
trigger turbulence and hence raise the viscosity up to turbulent values. From
the typical value of the microscopic Prandtl number, namely $10^{-6}$,
and that of critical Reynolds numbers, $\sim 10^3$, we can expect and
increase of Ekman number up to $2\, 10^{-15} \tv 2\, 10^{-12}$ and
hence a reduction of the circulation time down to $10^{11}\tv 10^8$ yrs
($10^8$ yrs is the Kelvin-Helmoltz time of our model). In the early phase
of evolution, when $\mu$-barriers are not strong enough, this time scale
may be further reduced by the presence of the Stewartson layer. However,
this layer is sensitive to the $\lambda$-number, which means sensitive
to the Prandtl number. Thus, if turbulence is strong enough to raise the
Prandtl number to order unity values, the Stewartson layer disappears. On
the other side if it is mild enough and increases the viscosity by
a factor $10^3$ only, the Stewartson layer has a thickness \od{2\,10^{-4}}
which reduces the circulation time further by a factor $5 10^3$; all in all,
mild turbulence and Stewarson layer may reduce the circulation time by a
factor $5 10^6$ leaving a time scale for partial mixing around $10^7$yrs. On
the other hand if turbulence is strong, circulation outside 
the thermal layer, which reaches the core, is $\sim
\PR^{-1}$ faster thanks to the increased viscosity, but $\lambda^{-1}$
slower because of stratification. However, the $\lambda^{-1}$ factor
only reduces the advection time, if turbulence is strong, elements may
diffuse just like momentum and the time scale is $T_{\rm visc}{\cal
P}=10^8$yrs.

The point raised here is, rather than the numbers, the role played by
the different hydrodynamical processes in controlling the transport of
element. Interestingly enough, we see that these processes may control
each other: a mild turbulence being helped by a strong Stewartson layer
yields a mixing time scale not much different from the one derived in the
strong turbulence case.  This leaves the possibility that mixing may
not be too sensitive to the hydrodynamical details. In particular, we do
not include angular momentum loss through a stellar wind
\cite[see][]{zahn92}; our result shows the possible mixing occuring in
the weak wind limit.

To end this section let us put some numbers on the differential rotation
induced by baroclinicity. The scalings give the following relation:

\[ \frac{\Delta\Omega}{\Omega} = 2\RO\, \delta \Omega\]
where $\delta \Omega=u_\varphi/s$. The numerical solutions show
that $\delta \Omega_{\rm pole} - \delta \Omega_{\rm eq} \sim 0.5$ when
$\PR\ll1$ and decreases to 0.01=\od{\lambda^{-1}} when $\PR\sim1$.

From these values we see that such a model predicts a rather important
differential rotation when the Prandtl number is small and a strong
reduction of it if some turbulence develops.

\subsection{Non-linear terms}

Discussing the results, we may wonder about the importance of the
non-linear terms that we neglected. They are $\RO\,(\vu\cdot\na\vu)$
and $\varepsilon\vu\cdot\na{\theta}$. Neglecting \od{E} terms,
the momentum nonlinear term is just $-\RO\,u^2_\varphi/s\ephi$, i.e. a
centrifugal correction to that of the background rotation. We thus
expect a slight modification of the baroclinicity but no qualitatively
important change.

Concerning the heat advection, the weakness of meridional circulation
\od{E}, plus the \od{\lambda} amplitude of $\theta$, make this nonlinear
term \od{\varepsilon\lambda E}, thus extremely small and really
negligible.

It hence turns out that nonlinear terms are either very small or not
bringing new physical phenomena in the steady solution. Taking into
account the initial assumption of the Boussinesq approximation, refining
the solution with these terms does not make sense.

\subsection{Back to the Boussinesq approximation}

{\bf
To conclude this discussion, we briefly come back on the use of the
Boussinesq approximation. One may indeed wonder whether in a more
realistic approach, taking into account density variations of the
background, our results would persist. The Boussinesq approximation
is indeed the consequence of two limits: the fluid velocity is small
compared to the sound velocity and the density scale height is large
compared to size of the container \cite[see][]{Rieu97}. The first
constraint is easily met: Mach numbers are small. The second one
is not because the density varies on a scale comparable to the size of the
star. Thus, the simplification of the equation of mass conservation from
$\Div\rho\vv=0$ into $\Div\vv=0$ is certainly abusive and quantitatively
our model should be taken with care.  On the qualitative side, however,
we note that density variations cannot modify the hydrodynamical features
like the sign of the differential rotation, the appearance of Stewartson
layer, etc. One may actually note that the use of the momentum $\rho\vv$
instead of $\vv$ would solve this problem, but then be inconsistent
with the neglect of the effects of centrifugal distorsion.  Thus, even
if at first sight the use of the \BA\ seems exagerated for a star, it
provides an interesting view of its hydrodynamics: it takes into account
all the force fields which control the flows and gives a physically
self-consistent model.
}

\subsection{Conclusions}

To end this paper we would like to stress the most interesting points of
this model, namely:

\begin{itemize}
\item The determination of the differential rotation in radiative
envelope as a function of the \BV\ frequency profile and its sensitivity
to the amplitude of a turbulent viscosity,
\item The demonstration that, when the \BV\ frequency varies on a scale
of the order of the radius, baroclinicity generates a fast rotating pole
and slow equator and the opposite in the case the envelope is
convective,
\item The determination of the meridian circulation, its shape and
amplitude, its sensitivity to Prandtl number and \BV\ frequency profile,
\item The appearance of Stewartson layers induced by the jump in
the mechanical properties of the fluid at the core-envelope boundary and the
sensitivity of this layer to the $\mu$-gradients or the Prandtl number,
\item The richness of the dynamics of these envelope and the
compensation, as far as mixing is concerned, which may result from
large-scale flows (Stewartson layer) and small-scale turbulence.
\end{itemize}

Naturally, such a model cannot pretend to describe actual stars at face
value. It should rather be viewed as a laboratory experiment aimed at
studying some aspects of the stellar dynamics. Its (relative) simplicity
indeed authorizes the study of hydrodynamical instabilities at global
scale through either linear analysis or full numerical simulations.
For instance, one may think to use it to study dynamo effect in radiative
envelope following \cite{BS05} and to study the effects of an additional
driving like the angular momentum loss generated by a wind.

The next step is to compute the same flow in the real, spheroidal,
geometry of rapidly rotating stars. Such realistic models will
determine the aspects of the dynamics of stellar envelopes which are
robust and can be studied with our Boussinesq model. They may also
suggest some intermediate (artificial) models where a baroclinic torque
is just plugged in a spherically symmetric stellar model for studying the
dynamical aspects which are not sensitive to the true spheroidal shape.

\begin{acknowledgements}
I would like to thank  Matthieu Castro, Boris Dintrans, Bernard Pichon
for providing me with various profiles of \BV\ frequency out of stellar
evolution codes; I am also indebted to François Lignières, Sylvie Vauclair
and Jean-Paul Zahn for many useful discussions. This work is part of
the ESTER project aimed at modelling stars in two dimensions; it is
supported by the Programme National de Physique Stellaire along
with the Action Spécifique pour la Simulation Numérique en Astrophysique.

This work was started while the author was hosted at Newton Institute
during the programme 'Magnetohydrodynamics of Stellar Interiors' which is
gratefully acknowledged.
\end{acknowledgements}

\bibliographystyle{aa}
\bibliography{../../biblio/bibnew}

\end{document}

%% file: com31.tex
\def\bbbc{{\mathchoice {\setbox0=\hbox{$\displaystyle\rm C$}\hbox{\hbox
to0pt{\kern0.4\wd0\vrule height0.9\ht0\hss}\box0}}
{\setbox0=\hbox{$\textstyle\rm C$}\hbox{\hbox
to0pt{\kern0.4\wd0\vrule height0.9\ht0\hss}\box0}}
{\setbox0=\hbox{$\scriptstyle\rm C$}\hbox{\hbox
to0pt{\kern0.4\wd0\vrule height0.9\ht0\hss}\box0}}
{\setbox0=\hbox{$\scriptscriptstyle\rm C$}\hbox{\hbox
to0pt{\kern0.4\wd0\vrule height0.9\ht0\hss}\box0}}}}
\def\bbbq{{\mathchoice {\setbox0=\hbox{$\displaystyle\rm
Q$}\hbox{\raise
0.15\ht0\hbox to0pt{\kern0.4\wd0\vrule height0.8\ht0\hss}\box0}}
{\setbox0=\hbox{$\textstyle\rm Q$}\hbox{\raise
0.15\ht0\hbox to0pt{\kern0.4\wd0\vrule height0.8\ht0\hss}\box0}}
{\setbox0=\hbox{$\scriptstyle\rm Q$}\hbox{\raise
0.15\ht0\hbox to0pt{\kern0.4\wd0\vrule height0.7\ht0\hss}\box0}}
{\setbox0=\hbox{$\scriptscriptstyle\rm Q$}\hbox{\raise
0.15\ht0\hbox to0pt{\kern0.4\wd0\vrule height0.7\ht0\hss}\box0}}}}
\def\bbbt{{\mathchoice {\setbox0=\hbox{$\displaystyle\rm
T$}\hbox{\hbox to0pt{\kern0.3\wd0\vrule height0.9\ht0\hss}\box0}}
{\setbox0=\hbox{$\textstyle\rm T$}\hbox{\hbox
to0pt{\kern0.3\wd0\vrule height0.9\ht0\hss}\box0}}
{\setbox0=\hbox{$\scriptstyle\rm T$}\hbox{\hbox
to0pt{\kern0.3\wd0\vrule height0.9\ht0\hss}\box0}}
{\setbox0=\hbox{$\scriptscriptstyle\rm T$}\hbox{\hbox
to0pt{\kern0.3\wd0\vrule height0.9\ht0\hss}\box0}}}}
\def\bbbz{{\mathchoice {\hbox{$\sf\textstyle Z\kern-0.4em Z$}}
{\hbox{$\sf\textstyle Z\kern-0.4em Z$}}
{\hbox{$\sf\scriptstyle Z\kern-0.3em Z$}}
{\hbox{$\sf\scriptscriptstyle Z\kern-0.2em Z$}}}}

\newcommand{\para}{{/\!/}}

\newcommand{\BA}{Boussinesq approximation}
\newcommand{\BV}{Brunt-V\"ais\"al\"a\ }
\newcommand{\BVF}{Brunt-V\"ais\"al\"a frequency}

\newcommand{\beq}{\begin{equation}}
\newcommand{\beqa}{\begin{eqnarray*}}
\newcommand{\beqan}{\begin{eqnarray}}
\newcommand{\greq}{\begin{equation}\left\{ \begin{array}{l}}
\newcommand{\egreq}{\end{array}\right. \end{equation}}
\newcommand{\nngreq}{\[\left\{ \begin{array}{l}}
\newcommand{\nnegreq}{\end{array}\right. \]}
\newcommand{\egreqn}[1]{\end{array}\right. \label{#1}\end{equation}}
\newcommand{\eeq}{\end{equation}} 
\newcommand{\eeqn}[1]{\label{#1}\end{equation}} 
\newcommand{\eeqa}{\end{eqnarray*}}
\newcommand{\eeqan}[1]{\label{#1}\end{eqnarray}}

\newcommand{\ssi}{ \Longleftrightarrow}

\newcommand{\tv}{ \rightarrow}

\newcommand{\lp}{ \left(}
\newcommand{\rp}{ \right)}
\newcommand{\lc}{ \left[}
\newcommand{\rc}{ \right]}
\renewcommand{\la}{ \left\{}
\newcommand{\ra}{ \right\}}
\newcommand{\khi}{\chi}
\newcommand{\llp}{\ell(\ell+1)}
\newcommand{\clm}{c^\ell_m}
\newcommand{\plm}{p^\ell_m}
\newcommand{\ulm}{u^\ell_m}
\newcommand{\vlm}{v^\ell_m}
\newcommand{\wlm}{w^\ell_m}
\newcommand{\thetalm}{\theta^\ell_m}
\newcommand{\YL}{Y^m_\ell}

\newcommand{\RL}{ \vec{R}^m_\ell }
\newcommand{\SL}{ \vec{S}^m_\ell }

\newcommand{\TL}{ \vec{T}^m_\ell }

\newcommand{\eps}{\varepsilon}

\renewcommand{\na}{ \vec{\nabla} }

\newcommand{\intvol}{ \int_{(V)}\! }
\newcommand{\cth}{ \cos\theta }

\newcommand{\sth}{ \sin\theta }

\newcommand{\FR}{\mbox{$ {\rm Fr} $}}

\newcommand{\PR}{\mbox{$ {\cal P} $}}

\newcommand{\RE}{\mbox{$ {\rm Re} $}}
\newcommand{\RI}{\mbox{$ {\rm Ri} $}}
\newcommand{\RO}{\mbox{$ {\rm Ro} $}}

\newcommand{\vu}{\vec{u}}

\newcommand{\er}{\vec{e}_r}

\newcommand{\ephi}{\vec{e}_\varphi}

\newcommand{\es}{\vec{e}_s}

\newcommand{\ez}{\vec{e}_z}

\newcommand{\vC}{\vec{C}}

\newcommand{\vg}{\vec{g}}

\newcommand{\vr}{\vec{r}}

\newcommand{\vv}{\vec{v}}

\newcommand{\vO}{\vec{\Omega}}

\newcommand{\vzero}{\vec{0}}

\newcommand{\tE}{\tilde{E}}

\newcommand{\tu}{\tilde{\vu}}
\newcommand{\bu}{\overline{u}}

\newcommand{\demi}{\frac{1}{2}}

\newcommand{\od}[1]{\mbox{${\cal O}(#1)$}}

\newcommand{\dr}[1]{\frac{\partial  #1}{\partial r}}

\newcommand{\ds}[1]{\frac{\partial  #1}{\partial s}}

\newcommand{\dnr}[1]{\frac{d  #1}{dr}}

\newcommand{\ddnr}[1]{\frac{d^2  #1}{dr^2}}

\newcommand{\dddnr}[1]{\frac{d^3  #1}{dr^3}}

\newcommand{\dtheta}[1]{\frac{\partial  #1}{\partial \theta}}

\newcommand{\drr}{\frac{\partial}{\partial r}}

\def\Div{\mathop{\hbox{div}}\nolimits}

\def\rot{\mathop{\hbox{\overrightarrow{\displaystyle\rm Rot}}}\nolimits}

\renewcommand{\rot}{\vec{\nabla}\times}

\renewcommand{\Div}{\vec{\nabla}\cdot}

\newcommand{\eq}[1]{(\ref{#1})}

\newcommand{\infapp}{\raisebox{-.7ex}{$\stackrel{<}{\,\sim\,}$}}

